# Navigating the landscape of COVID-19 research through literature analysis: A bird's eye view.

Lana Yeganova [§], Rezarta Islamaj [§], Qingyu Chen, Robert Leaman, Alexis Allot, Chin-Hsuan Wei, Donald C. Comeau, Won Kim, Yifan Peng, W. John Wilbur, Zhiyong Lu[*].

National Center for Biotechnology Information (NCBI), NLM, NIH, Bethesda, MD USA

{Lana.Yeganova, Rezarta.Islamaj, Qingyu.Chen, Robert.Leaman, Alexis.Allot, Chih-Hsuan.Wei, Donald.Comeau, Won.Kim, Yifan.Peng, John.Wilbur, Zhiyong.Lu}@nih.gov

## ABSTRACT

Timely access to accurate scientific literature in the battle with the ongoing COVID-19 pandemic is critical. An unprecedented public health risk, this pandemic has motivated research towards understanding the disease in general, identifying drugs to treat the disease, developing potential vaccines, etc. This has given rise to a rapidly growing body of literature that doubles in number of publications every 20 days as of May 2020. Providing medical professionals with means to quickly analyze the literature and discover growing areas of knowledge is necessary for addressing their question and information needs.

In this study we analyze the LitCovid collection, 13,369 COVID-19 related articles found and curated from PubMed as of May 15[th], 2020 with the purpose of examining the landscape of literature data and presenting it in a format that facilitates information navigation and understanding. We do that by applying state-of-the-art named entity recognition, classification, clustering and other natural language processing (NLP) techniques. Applying named entity recognition tools, we capture relevant bioentities (such as diseases, internal body organs, etc.) and assess the strength of their relationship with COVID-19 by the extent they are discussed in the corpus. We also collect a variety of symptoms and co-morbidities discussed in reference to COVID-19. Our clustering algorithm identifies topics represented by groups of related terms, and computes clusters corresponding to documents associated with the topic terms. Among the topics we observe several that persist through the duration of multiple weeks and have numerous associated documents, as well several that appear as emerging topics with fewer documents. All the tools and data are publicly available, and this framework can be applied to any literature collection. Taken together, these analyses produce a comprehensive, synthesized view of COVID-19 research to facilitate knowledge discovery from literature.

## 1 Introduction

Keeping up to date with COVID-19 related research is increasingly challenging as the number of publications on the topic doubles every 20 days. Since the beginning of the pandemic, the natural language processing community has responded to the needs of domain experts by applying state-of-the-art AI and NLP methods to improve information understanding, comprehension, and discoverability. These efforts include: LitCovid (https://www.ncbi.nlm.nih.gov/research/coronavirus/), a daily updated website curating and providing access to the most comprehensive and informative COVID-19 publications in PubMed (Chen, Allot et al. 2020); CORD-19, a weekly updated dataset of articles on COVID-19 and other coronaviruses such as MERS and SARS (https://www.semanticscholar.org/cord19); https://www.semanticscholar.org/cord19); Neural Covidex (covidex.ai), a neural network based question answering system.

Due to the rapid growth of the COVID-19 related literature, there is an unmet urgency for helping the biomedical and clinical community keep afloat in a sea of literature(Brainard 2020). At the same time this presents an extraordinary opportunity for research in the field of natural language processing to be of critical importance. There are already examples of recent studies examining this growing body of literature, including comprehensive named entity recognition (Wang, Song et al. 2020), automatic textual evidence extraction (Wang, Liu et al. 2020), topic analysis and visualization (Bras, Gharavi et al. 2020), among others. In our approach we combine several directions, including NER and topic analysis and visualization, to paint a comprehensive picture and juxtapose findings from complementary analyses.

To this end, we present the entire COVID-19 literature, through automated text analytics, in a format that facilitates understanding and information navigation to help researchers and medical professionals grasp the literature landscape of the disease. We analyze the collection of PubMed articles made available through the LitCovid website; we will refer to these articles as the LitCovid collection. To analyze the articles in the LitCovid collection we make an extensive use of available NLP tools, including named entity recognition (NER) methods, supervised multiclass classification strategies, and unsupervised clustering algorithms. Using the unsupervised topic analysis and clustering approach (Islamaj, Yeganova et al. 2020) we discover meaningful topics in the dataset, including both major topics rich in terms and with many relevant articles, as well as smaller clusters representing emerging topics or trends.

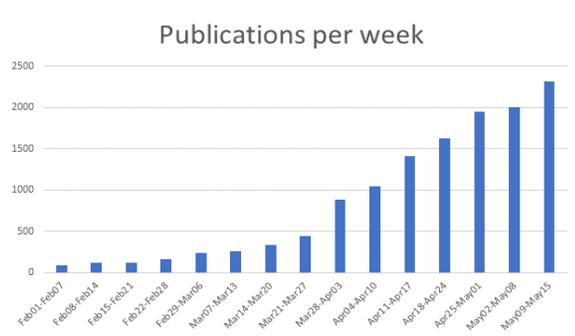

Figure 1 Number of publications available in PubMed in weekly increments from February1 to May 15, 2020.

## 1 LitCovid collection and characteristics

Our LitCovid dataset includes 13,369 PubMed articles as of May 15, 2020. Visualizing the characteristics of the dataset is a useful step in exploratory analysis of the literature collection.

**Temporal characteristics.** Projecting articles on a timeline allows one to observe spikes in the events relevant to the topic of interest. For the COVID-19 pandemic we observe an exponential growth rate of publications from January 2020 to the present, May 15th, 2020 (Figure 1). Literature collections could experience growth spikes due to various reasons such as a pandemic, a scientific breakthrough, increase/decrease in funding, etc. In the case of COVID-19, the expansion is not unexpected due to the severity of the outbreak and the determination of scientists worldwide to engage the problem.

**Semantic classification of articles in LitCovid collection.** Since the beginning, the articles in the LitCovid collection have been categorized into eight broad semantic categories. Specifically, topics were assigned by examining article titles, abstracts and, as needed, the full article text. The first batch of documents was annotated manually. Publications in the LitCovid collection are distributed among the topics as follows: Prevention (5,466 articles), Treatment (2,802 articles), Diagnosis (1,947 articles), Mechanism (1,487 articles), General Information (928 articles), Case Reports (832 articles), Transmission (673), and Epidemic Forecasting (227 articles). 1,843 articles were not identified with any category.

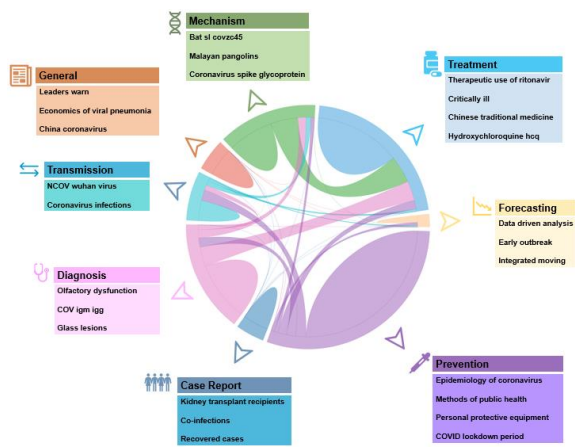

Figure 2 Distribution of COVID-19 articles by Semantic Categories: Prevention, Treatment, Mechanism, Diagnosis, Epidemiologic Forecasting, Case Report, General Information.

Furthermore, PubMed articles can be attributed to more than one category, resulting in 2,002 articles attributed to two categories, and 375 articles to three or more categories. The most frequent co-occuring categories are Mechanism and Treatment (888 pmids), Diagnosis and Treatment (641 pmids), and Transmission and Prevention(344 pmids), as shown in Figure 2.

## 2 Biomedical Entities in LitCovid collection

We identified most-occurring biomedical concepts in the COVID-19 literature by extracting the concept annotations in PubTator for the LitCovid collection. PubTator is a repository of biomedical concept annotations (Wei, Allot et al. 2019), powered by state-of-the-art entity taggers to identify six concept types: GNormPlus (Wei, Kao et al. 2015) for genes, TaggerOne (Leaman and Lu 2016) for diseases and cell lines, tmVar (Wei, Phan et al. 2018) for genetic variants, SR4GN (Wei, Kao et al. 2012) for species and a BlueBERT-based tagger (Peng, Yan et al. 2019) for chemicals. We used the PubTator annotations from 5/28/2020 (https://github.com/ncbi-nlp/PubTator-Covid19). We calculated the article mention counts for each entity, that is, the number of articles that mention the entity at least once, including synonyms (e.g. remdesivir and GS-5734 refer to the same drug). These results allow us to observe the entities mentioned and analyze their significance over time.

The five drugs most frequently mentioned are hydroxychloroquine, chloroquine, remdesivir, tocilizumab, and lopinavir/ritonavir drug combination. Chloroquine and hydroxychloroquine combined are mentioned extensively, with hydroxychloroquine dominating over the course of several weeks, starting from the week of February 29, 2020. We find many studies targeted at understanding the adverse effects of hydroxychloroquine. Remdesivir and the lopinavir/ritonavir drug combination are the earliest drugs discussed in the corpus.

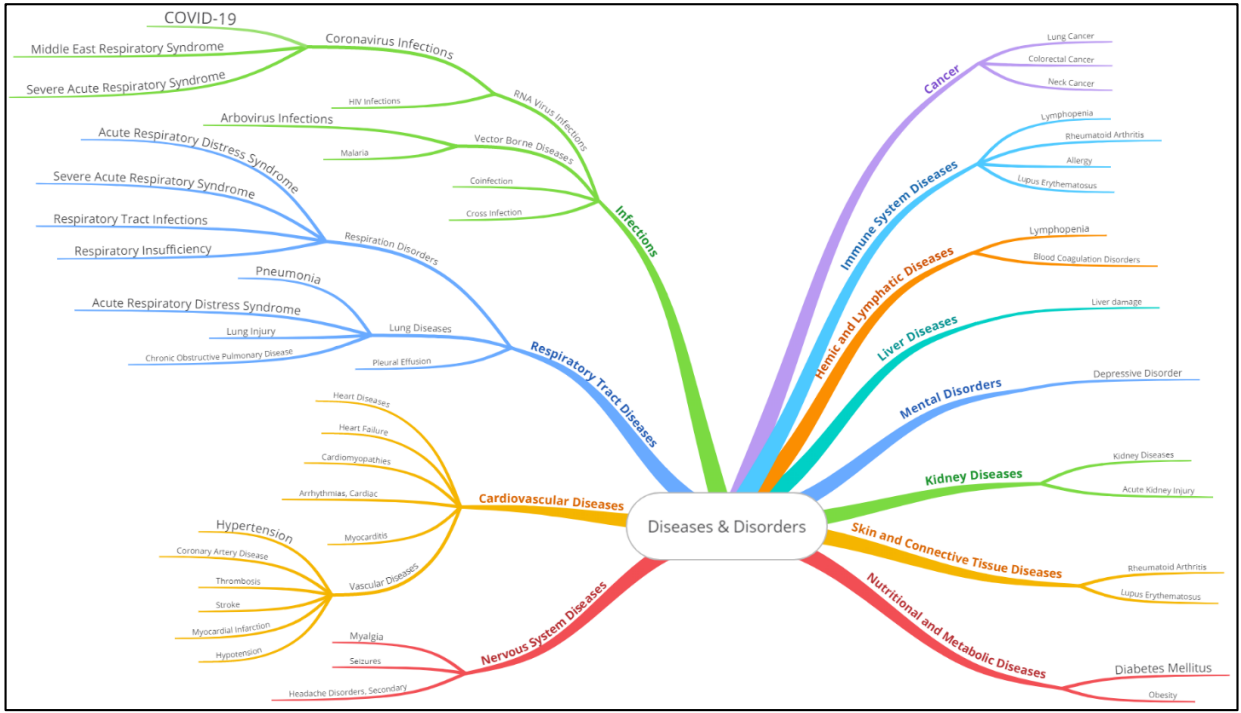

Figure 3 The most relevant diseases in hierarchy structure. The font size of the disease reflects the frequency.

Table 1 The most frequent genes, cell lines and variants in the LitCovid collection, with their frequencies (in parentheses).

| Top Concepts | Human gene | SARS-CoV-2 gene | Human variant | | Covid19 variant | | CellLine |
|---|---|---|---|---|---|---|---|
| | | | Variant | Corresponding Gene | Variant | Corresponding gene | |
| 1 | ACE2 (1127) | Spike (236) | K479N (3) | ACE2 | D614G (7) | Spike | Vero E6 (99) |
| 2 | CRP (680) | Orf1a/b (193) | S487T (3) | ACE2 | V553L (5) | Orf1a/b | HeLa (40) |
| 3 | Il-6 (528) | Nucleocapsid (166) | rs143936283 (E329G) (2) | ACE2 | L3606F (5) | Orf1a/b | MRC5 (37) |
| 4 | ACE (373) | Envelope (76) | rs2285666 (8790G>A) (2) | ACE2 | G251V (5) | ORF3a | Hussain SY (30) |
| 5 | TNF-a (318) | Membrane (72) | rs755691167 (K68E) (2) | ACE2 | F476L (4) | Orf1a/b | Huh7 (29) |

The diseases most frequently mentioned are illustrated in Figure 3 and categorized into 12 high-level groups. Most of the frequently mentioned diseases (other than COVID-19 itself) are respiratory disorders (e.g. respiratory insufficiency, pneumonia and lung injury) that are common effects of COVID-19, or related viral infections (e.g. SARS, MERS and arbovirus infections). The remainder include diseases that raise the risk of developing severe complications – primarily cardiovascular diseases (e.g., hypertension), metabolic diseases (e.g., diabetes) – and pathologies caused by severe COVID-19 – renal failure, liver damage, and complications associated with thrombosis (blood clots) such as stroke. Less frequently mentioned are diseases that require special management during COVID-19 (especially cancers, though also diabetes and cardiovascular diseases), associated conditions used to predict severity (such as lymphopenia) and mental health conditions caused by the pandemic. Rheumatoid arthritis and lupus are mentioned primarily because hydroxychloroquine is an established treatment for both.

Table 1 lists the five most frequently mentioned genes, genetic variants and cell lines. Analysis of these entities reflects both the current knowledge regarding SARS-CoV-2 infection and current areas of investigation. ACE2 (angiotensin-converting enzyme 2) is the receptor SARS-CoV-2 uses to enter cells (Zhou, Yang et al. 2020); it is the human gene most frequently mentioned, and the only gene associated with any of the five most frequent human genetic variations. Levels of CRP (c-reactive protein), IL-6 (interleukein-6), and TNF-alpha (tumor necrosis factor alpha) are used as markers of SARS-CoV-2 infection severity (Ni, Tian et al.

Table 2 Article mentions of selected symptoms in the LitCovid collection. Symptoms are from the MeSH C23 tree and can be either a result of the disease or a pre-existing condition. We distinguish frequent, moderate and rare symptoms.

| Article Mentions | Symptom |
|---|---|
| 559 | Fever |
| 390 | Cough |
| 174 | Inflammation |
| 153 | Fatigue |
| 113 | Diarrhea |
| <100 | Headache, Olfaction Disorders, Infarction, Hypoxia, Asymptomatic Infection, Neurologic Manifestations |
| <10 | Fetal Distress, Obesity, Morbid, Taste Disorders, Overweight, Hyperventilation, Death, Sudden, Weight Loss, Hearing Loss, Lethargy |

2020). The use of ACE (angiotensin-converting enzyme) inhibitors, normally used to treat hypertension, has been investigated as potentially increasing the probability of severe disease (Diaz 2020) Four of the most frequently mentioned SARS-CoV-2 genes refer to structural proteins: spike, nucleocapsid, envelope and membrane (Khailany, Safdar et al. 2020) spike is the protein used to bind ACE2 to gain entry to the cell (Lan, Ge et al. 2020) and it is the gene associated with the most frequent SARS-CoV-2 genetic variant. ORFa/b refers to an open reading frame; it is under investigation for clinically significant genetic variants; accordingly, it is the SARS-CoV-2 gene associated with the majority of frequent SARS-CoV-2 genetic variants. The most frequently mentioned cell line is Vero E6, which is frequently used as host cells in viral studies, as is HuhH7. HeLa and MRC5 are common cell lines. Considering the most frequently mentioned entities as a whole, it becomes apparent that the current primary focus of genetic research related to COVID-19 is how SARS-CoV-2 functions – and by extension, what interventions might interrupt critical functions – with a secondary focus on clinical markers of disease severity and the genetic variants that influence it.

## 3 Symptoms and findings in LitCovid collection.

One of the most frequently asked questions in the media is "What are the symptoms of this disease?". MeSH (Medical Subject Headings) is the controlled vocabulary thesaurus at the National Library of Medicine used for indexing articles. Article mention counts were collected using the Pathological Conditions, Signs and Symptoms branch (C23) of MeSH as a dictionary of symptoms. Counts for MeSH entry terms (synonyms) were added to the count of each MeSH Heading. Mentions are detected in 7,975 of the articles in the LitCovid collection. Table 2 illustrates examples of symptoms ranging from frequent through moderate and rare. It is important to note that some of these symptoms are caused by COVID-19. Others are related to underlying conditions that make patients more susceptible or increase the chance of a serious case. Comparison of the LitCovid collection symptoms with the symptom mentions of a similar disease, SARS, allows detection of symptoms specific for COVID-19. For the SARS collection we chose all 9,085 PubMed articles mentioning SARS through June 2019. Even though this set includes years of research, it is smaller than the LitCovid set. This reflects the volume and intensity of COVID-19 research. We computed 95% confidence intervals around the frequencies of symptoms appearing in both sets. If the intervals did not overlap, the frequency differences are significant. Table 3 reports selected symptoms that are unique to each corpus as well as some that have significantly different frequencies.

Table 3 Comparison of symptoms in the COVID and SARS literature sets. Selected symptoms that either appear in only one set, or appear significantly more frequently in one set. Symptoms are from the MeSH C23 tree and can result from the disease or a pre-existing condition.

| COVID Only |
|---|
| Olfaction Disorders // Myocardial Infarction |
| Abdominal Pain // Dizziness |
| **SARS Only** |
| Carcinogenesis // Chromosome Aberrations |
| **COVID Significantly More Frequent** |
| Death // Inflammation // Diarrhea |
| Chronic Disease // Recurrence |
| **SARS Significantly More Frequent** |
| Osteonecrosis |
| **No Significant Difference in Mentions** |
| Communicable Diseases // Sneezing // Convalescence |

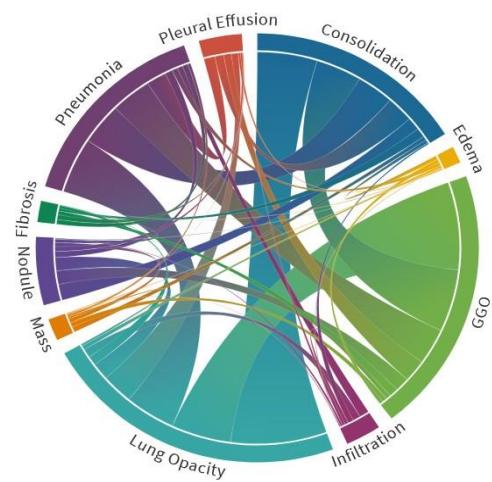

Figure 4 The most frequent pathological findings in chest radiograph and computed tomography studies

Further, we examined the common pathological findings in chest x-ray and computed tomography (CT) studies reported in the LitCovid collection. The findings are based on common thoracic disease pathology types, which are expanded from NIH Chest X-ray 14 labels (Wang et al., 2017). Specifically, we used our previous tool, NegBio, to extract these from the figure caption of chest x-ray and CT images in the full-length PMC OA articles in the LitCovid collection (Peng et al., 2017). Figure 4 illustrates the passage-level co-mentions of the top ten findings: *Ground Glass Opacities (GGO), Consolidation, Lung Opacity, Infiltration, Pneumonia, Pleural Effusion, Edema, Mass, Nodule,* and *Fibrosis*. The most notable mention co-occurrence is between *Lung Opacity* and *Consolidation*, appearing in 139 articles.

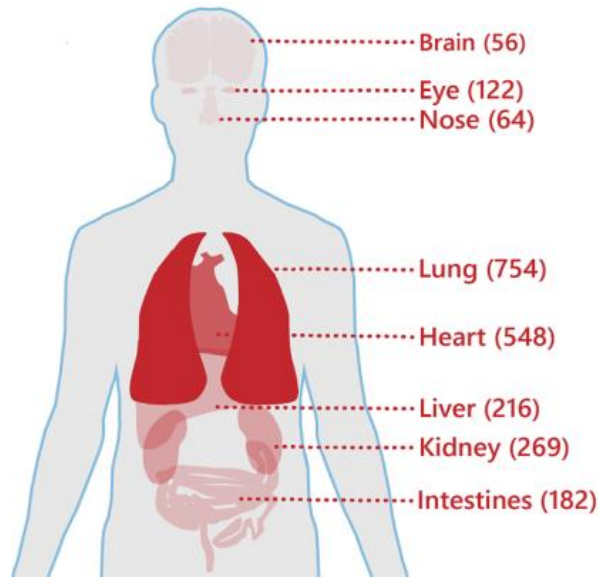

Figure 5 Human organ intensity heat map illustrating the frequency of mentions in the LitCovid collection.

## 4 Human organs affected by SARS-CoV-2.

COVID-19 has been linked to causing multi-system organ failure. In this experiment, we wanted to illustrate and summarize the research literature as localized by the human organ mentioned. Organ mentions are extracted using the en_ner_bionlp13cg_md model from SciSpacy (Neumann, King et al. 2019), and entities labelled {'ORGAN', 'ANATOMICAL_SYSTEM', 'IMMATERIAL_ANATOMICAL', 'DEVELOPING_ANATOMICAL_STRUCTURE'} mapped to UMLS terms using the UMLS normalizer from SciSpacy, with a similarity threshold of 0.9. Matching terms were ordered by length, and the shortest was used. Figure 5 is a human organ intensity map illustrating the number of article mentions of the top 10 organs.

## 5 Clustering the LitCovid literature collection

Clustering methods can be useful for automatically grouping documents into clusters and/or terms into meaningful topics, improving human comprehension of a document collections. In this work we rely on the Probabilistic Distributional Clustering (PDC) algorithm (Islamaj, Yeganova et al. 2020) to generate topics and associated document clusters for the LitCovid literature collection. Given a document collection, the algorithm computes disjoint term sets representing topics in the collection. The algorithm relies on the analysis of words and two-word phrases and their frequencies and co-occurrences in the free text. It partitions the set of terms appearing in the collection of documents into disjoint groups of related terms in an unsupervised way. This algorithm is useful in that it does not require a user-provided input for the number of topics/clusters and does not require any pre-existing knowledge or vocabulary for a given body of literature, which is ideal for applying to a newly emerged corpus.

Using this algorithm, we discover meaningful topics that arise from the dataset, including both large-size topics defined by many terms and retrieving many relevant articles, as well as smaller peaks in data representing emerging topics or trends. For each topic, we generate titles from the top scoring terms using intuitive rule-based techniques. Given a topic we also score all documents in the collection by computing a topic-document score based on presence of the topic terms in the document. For clarity, we will refer to the term sets as topics, and groups of retrieved relevant pmids as clusters.

The PDC-identified clusters on the LitCovid collection provide a comprehensive overview of the disease, as reconstructed from the literature, as shown in Figure 6. These clusters correspond to general and specific topics in dealing with the new disease. We observe clusters on treatment options, and case reports, patient characteristics and categories. There are clusters of documents describing both general topics, as well as specific ones associated with different ongoing conditions, such as cardiovascular diseases, ongoing cancer treatments etc. We see reports and analysis on different drugs, studies on the disease mechanism, testing options and other observed challenges and lessons learned. In addition, there are clusters of documents on the quality of life and issues of mental health in such a situation where life has been disrupted for many people.

### 5.1 Significant topics in the LitCovid collection.

The PDC algorithm applied to the LitCovid collection resulted in 289 clusters with 10 terms or more (we use the threshold of ten terms following the choice of parameters in (Islamaj, Yeganova et al. 2020)). Topics containing less than 10 terms are usually very specific and are considered separately in Section 3.2.

The significance of a cluster generated by the PDC algorithm is naturally correlated with the number of terms in the topic. The most term-rich topics produced by the algorithm are: "*severe acute respiratory syndrome*", "*personal protective equipment*", "*angiotensin converting enzyme*". A large term size generally signifies a set of vocabulary terms that are closely related and are used together in the literature. Large topics naturally retrieve more documents because they contain more terms. However, the results also contain many documents returned for topics with a small number of terms, describing highly specific aspects of a disease. This is expected when we consider the highly specific terms that are discussed with the COVID-19 phenomena, which scientists are still trying to understand. For example, highly specific term groups that retrieve many documents are: "*cytokine storm*", "*lopinavir/ritonavir*", "*critically ill patients*", "*ethical resource allocation*", etc. Here we discuss dominant clusters for the LitCovid collection, both in relation to the term size, as well as in relation to the number of documents they retrieve.

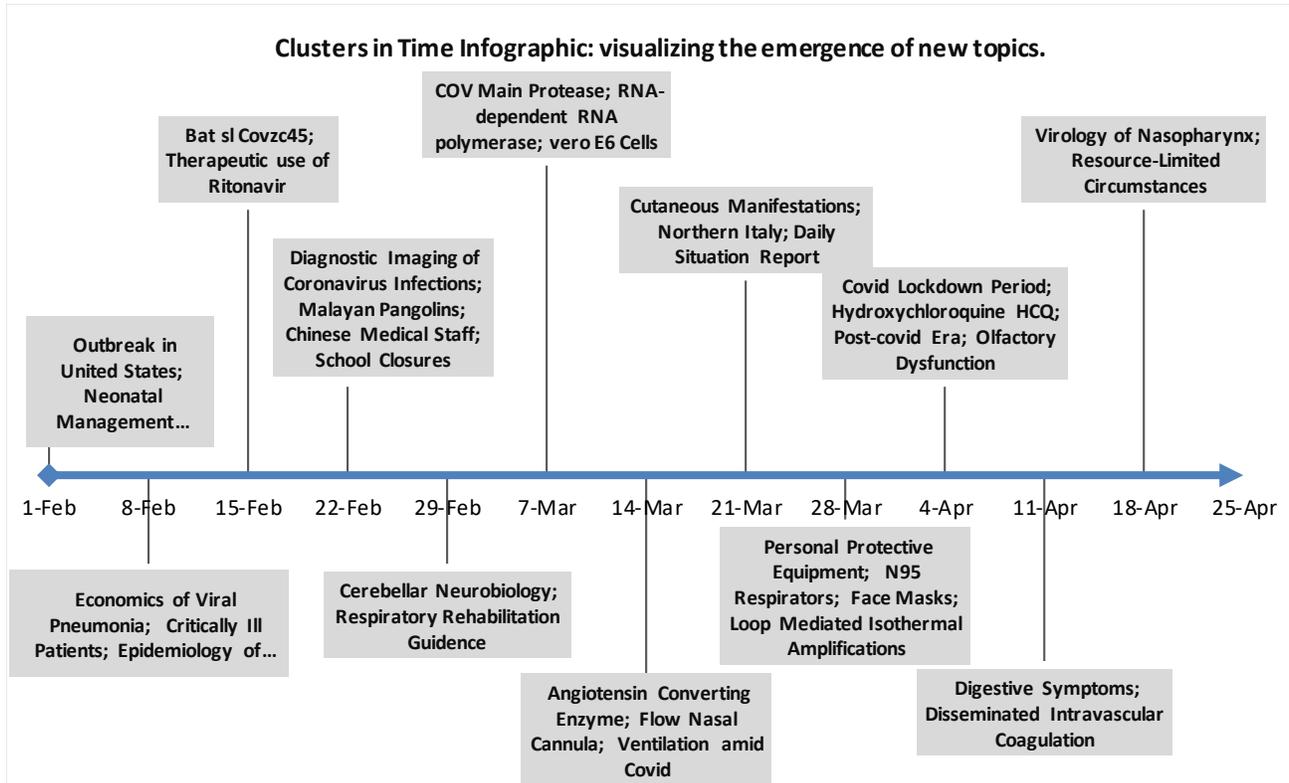

Figure 6 Computed clusters are illustrated along the timeline of the 100-day period starting Feb 1, 2020. Each point on the x-axis represents a week starting that day.

**Severe Acute Respiratory Syndrome.** The topic contains 290 unique terms and retrieves 295 pmids. Documents in this cluster are focused on the terms:

{ respiratory // sars // acute respiratory // cov // sars cov // syndrome // severe // acute // severe acute // respiratory syndrome}.

Top scoring articles identified to be relevant to this cluster are very much related to each other and to the topic terms:

- ```
  Epidemiology, virology, and clinical features of
  severe acute respiratory syndrome -coronavirus-2
  (SARS-CoV-2; Coronavirus Disease-19).
  ```
- ```
  Severe acute respiratory syndrome coronavirus 2
  (SARS-CoV-2) and coronavirus disease-2019 (COVID-
  19): The epidemic and the challenges.
  ```
- ```
  Asymptomatic carrier state, acute respiratory
  disease, and pneumonia due to severe acute
  respiratory syndrome coronavirus 2 (SARS-CoV-2):
  Facts and myths.
  ```
- ```
  The origin, transmission and clinical therapies
  on coronavirus disease 2019 (COVID-19) outbreak -
  an update on the status.
  ```
- ```
  The Coronavirus Pandemic: What Does the Evidence
  Show?
  ```

**Angiotensin Converting Enzyme.** The topic is focused on terms related to angiotensin converting enzyme and understanding of the disease mechanism. This topic is one of the most dominant topics of the corpus and consists of 115 unique terms and 278 relevant articles. Ten top scoring terms in the topic and relevant retrieved documents are:

{ angiotensin // converting enzyme // converting // angiotensin // converting // ace2 // renin // renin angiotensin // enzyme ace2 // blockers }

- ```
  Hypothesis: angiotensin-converting     enzyme
  inhibitors and angiotensin receptor blockers may
  increase the risk of severe COVID-19.
  ```
- ```
  ACE2, COVID-19, and ACE Inhibitor and ARB Use
  during the Pandemic: The Pediatric Perspective.
  ```
- ```
  Renin-Angiotensin System Blockers and the COVID-
  19 Pandemic: At Present There Is No Evidence to
  Abandon Renin-Angiotensin System Blockers.
  ```

**Pediatric Covid.** What about the pediatric patients? What are the special considerations, and how should the dosages be adjusted? This topic builds on the unique terms listed below, and retrieves 80 articles:

{ pediatric // pediatric covid // pediatric cases // overwhelm available // sras cov // total infected // preparedness efforts // sras // cumulative numbers }

- ```
  Pediatric Airway Management in COVID-19 patients
  - Consensus Guidelines from the Society for
  Pediatric Anesthesia's Pediatric Difficult
  Intubation Collaborative and the Canadian
  Pediatric Anesthesia Society.
  ```

- Brief report: International perspectives on the pediatric COVID-19 experience.
- COVID-19 in Children in the United States: Intensive Care Admissions, Estimated Total Infected, and Projected Numbers of Severe Pediatric Cases in 2020.
- Pediatric Airway Management in COVID-19 patients - Consensus Guidelines from the Society for Pediatric Anesthesia's Pediatric Difficult Intubation Collaborative and the Canadian Pediatric Anesthesia Society.
- Special considerations for the management of COVID-19 pediatric patients in the operating room and pediatric intensive care unit in a tertiary hospital in Singapore.
- Chloroquine dosing recommendations for pediatric COVID-19 supported by modeling and simulation.

These examples illustrate the value of the computed topics and related document clusters for both the broad view of the literature, as well as ability to navigate to and explore a document cluster of interest.

For the illustration in Figure 6, we have selected a subset of significant topics containing more than twenty terms and computed the onset of the topic from the earliest publication date among the most relevant documents. A week when the earliest and most relevant retrieved documents date to is considered as a topic onset. The topics in Figure 6 are organized by weeks. By studying the topics one can reconstruct the focus of published COVID-19 literature as a function of time. Some examples of early clusters are "*outbreak in United States*" and "*air travel*". As the outbreak evolves, clusters such as "*Diagnostic Imaging of Coronavirus Infection*" and "*Therapeutic use of Ritonavir*" start to emerge. In late March, topics "*Personal Protective Equipment*", "*N95 respirators*", and "Face masks" dominate the space. At the same time the community is actively exploring the "Hydroxychloroquine HCQ" treatment, and later we have clusters discussing the side effects of the treatment with HCQ.

## 5.2   Specific topics in the LitCovid collection.

Specific topics, as discussed above, consist of few terms (less than 10). However, this does not necessarily mean that they contain a small number of documents. As seen in Figure 7, clusters of topics *"Critically Ill Patients"*, or *"Lopinavir/Ritonavir"*, contain more than 50 PMIDs and they are spread throughout the 100 days. This shows that such topics remain of significant interest to researchers, and they are active research areas. However, most specific topics consist of a small set of very closely related documents. In this figure we illustrate several such specific clusters grouped into these larger categories:

1) *COVID-19 and Other Diseases*. In this category we illustrate specific cluster of documents discussing issues of patients diagnosed with COVID-19, who simultaneously suffer other conditions such as: *Fulminant Myocarditis, Acute Pulmonary*

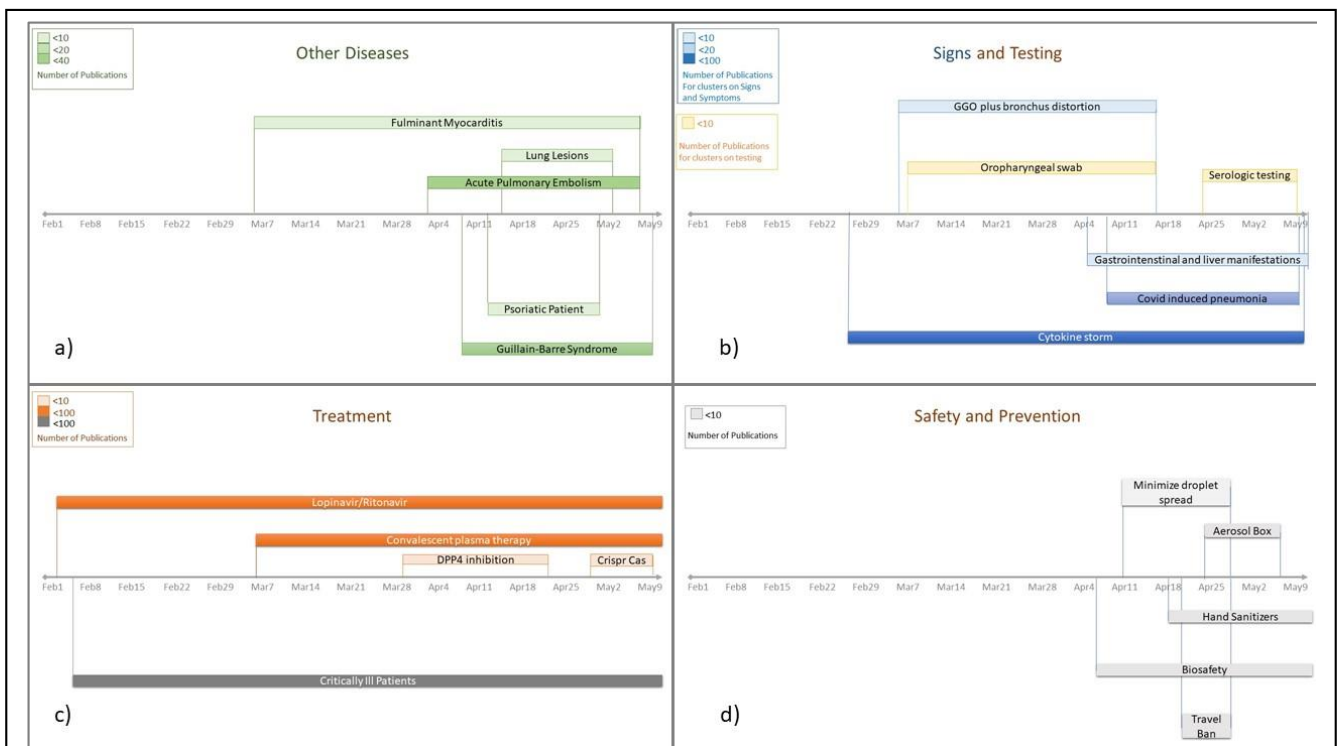

Figure 7 Small specific topics of LitCovid collection illustrated along the timeline. The figure illustrates small topics defined by few topic terms and the temporal spread of the document clusters: COVID-19 and a) other diseases, b) the diagnostic signs and testing procedures, c) different treatment options and their studies, and d) safety and prevention. The spread of the cluster reflects the publication dates of the documents collected in a cluster. The depth of color signifies the number of documents.

*Embolism, Lung Lesions, Guillain Barre Syndrome,* and *Psoriasis.*

2) *COVID-19 Signs and Testing*. In this category we have a specific topic, with many documents titled "*Cytokine Storm*". This active research area highlights the fact that this virus causes some patients to show a heightened immune response compared to others, and the factors that make some people more susceptible than others is still unknown. Other clusters in this category are: *GGO plus bronchus distortion, Gastrointestinal and liver manifestations,* and *COVID-induced pneumonia*. Some small clusters also reflect research on testing mechanisms.

3) *COVID-19 treatment*. This category shows an emerging set of papers on *Crispr-Cas* and research on using it for treatment purposes, and interesting questions related to the *DPP4 inhibitors* and the effect they could have on some patients with COVID-19, etc.

4) *Safety and Prevention*. Here we find a selection of small clusters discussing topics such as: *Hand sanitizer, Aerosol Box*, how to *Minimize droplet spread*, *Biosafety* and the impact of *Travel ban*.

In general, specific topics contain a smaller number of documents when compared to the most significant topics. The number of documents in these clusters is small because, for a document to be assigned to a cluster, it gets weighted with all the topic terms, and the combined weight should be above a certain threshold. What we see is that these could be highly specific areas of interest to physicians and researchers and that they discuss different aspects of COVID-19.

## 5.3 Associating clusters with the main categories.

By considering labels of the highest scoring documents retrieved for each cluster, we can associate every clusters with the category where most of its highest scoring documents fit. We collect all labels assigned to the top scoring five documents and associate a cluster with the most frequent label in that set. Here we refer the reader back to Figure 2, which illustrates some of these clusters and their mapping to general categories. For example, "*therapeutic use of ritonavir*" is assigned to *treatment*, "*reported olfactory dysfunction*" and "*glass lesions*" are attributed to diagnosis class. As documents are frequently labeled with more than one class, so are the clusters. Understandably, we observe clusters that frequently discuss *treatment* and *mechanism* together, and *prevention* and *transmission*. For example, "*angiotensin converting enzyme*" is attributed to the *mechanism* class but has a considerable number of documents that are assigned to both *mechanism* and *treatment*. The heatmap in Figure 8 illustrates this phenomenon by reflecting the percentage of clusters assigned to a topic with significant number of documents that are also assigned to the next most relevant class.

# 6 Conclusions

In this study we analyzed the LitCovid collection, 13,369 COVID-19 related articles (as of May 15, 2020). The purpose of this study is to provide a convenient and visual access for researchers and

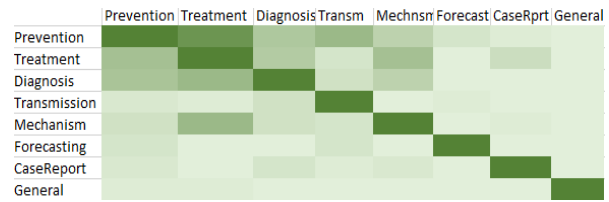

Figure 8 Heatmap represents the distribution of computed clusters across the eight main categories.

others to navigate the landscape of the COVID-19 literature. We present a bird eye's view of the literature as it spans broad categories and more specific topics in time. We identify the most frequently mentioned diseases, the most frequently mentioned organs, and both the most frequent and the most unusual symptoms, which we also compare with the published literature on SARS. In addition to the state-of-the-art named entity recognition tools, we also applied state of the art clustering methods. We computed topics represented by groups of related terms and corresponding clusters of documents associated with the topic terms. Among the topics we observe several that persist through the duration of multiple weeks and have numerous associated documents, while others appear as emerging topics with fewer documents. We further classify each of these topics into the major categories for easy information comprehension. This work is limited by the accuracy of the different tools that are used, as well as the human efforts required in the analysis. Our goal is to have this analysis available on the LitCovid website to benefit Covid-19 researchers (https://www.ncbi.nlm.nih.gov/research/coronavirus/). All the tools and data are publicly available, and this framework can be applied to any literature collection. For the Covid-19 research, taken together, these analyses produce a comprehensive, and synthesized view to facilitate knowledge discovery from literature.

## Funding

This work was supported by the Intramural Research Program of the National Library of Medicine, National Institutes of Health.